\documentclass[twocolumn,showpacs,preprintnumbers,amsmath,amssymb,prb]{revtex4}

\usepackage[pctex32]{epsfig}    
\usepackage{dcolumn}
\usepackage{bm}

\def\etal{{\em et al.\/}}
\def\kp{${\bf k}\cdot{\bf p}$}
\def\eva3{$eV$$\cdot$\AA$^3$}               
\def\vs{{\em vs.\/}}                    
\def\*#1*/{}                        
\def\iiiv{{III-V}}
\def\iiivs{{\iiiv s}}

\def\algaas{Al$_x$Ga$_{1-x}$As}
\def\ee #1{$\times 10^{#1}$}                     
\def\rsqbkt{$\rceil$\hspace{-5.1pt}$\rfloor$}   
\def\deg{\nobreak\hbox{\hskip 2.0 truept\hbox{$^\circ$}}}
\def\ie{{\em i.e.\/}}
\def\ML{\penalty 10000\hbox{${\,}$ML}}
\def\tr{\text{ tr}}

\newcommand{\ket}[1]{|#1\rangle}

\newcommand{\melem}[3]{\left\langle #1 \left| #2 \right| #3 \right\rangle }
\newcommand{\refeq}[1]{Eq.~(\ref{#1})}
\newcommand{\refeqs}[2]{Eqs.~(\ref{#1})-(\ref{#2})}

\newcommand {\NewBibliography}{%
  \bibliographystyle {apsrev}
  \bibliography{../../../papers/bibtex/complete}
}

\begin{document}

\title{Bulk Inversion Asymmetry effects on the band structure of zincblende heterostructures in an 8-band Effective Mass Approximation model}

\author{X. Cartoix\`a}
\affiliation{%
T. J. Watson, Sr., Laboratories of Applied Physics 128-95\\
California Institute of Technology\\
Pasadena, CA 91125, USA
}%

\author{D. Z.-Y. Ting}%
 \affiliation{
Jet Propulsion Laboratory\\
California Institute of Technology\\
Pasadena, CA 91109, USA
}%

\author{T. C. McGill}
 \email{tcm@ssdp.caltech.edu}
\affiliation{
T. J. Watson, Sr., Laboratories of Applied Physics 128-95\\
California Institute of Technology\\
Pasadena, CA 91125, USA
}%

\date{\today}

\begin{abstract}
We have developed an 8-band Effective Mass Approximation model that describes the zero field spin splitting in the band structure of zincblende heterostructures due to bulk inversion asymmetry (BIA). We have verified that our finite difference Hamiltonian transforms in almost all situations according to the true $D_{2d}$ or $C_{2v}$ symmetry of [001] heterostructures.  This makes it a computationally efficient tool for the accurate description of the band structure of heterostructures for spintronics. We first compute the band structure for an AlSb/GaSb/AlSb quantum well (QW), which presents only BIA, and delineate its effects. We then use our model to find the band structure of an AlSb/InAs/GaSb/AlSb QW and the relative contribution of structural and bulk inversion asymmetry to the spin splitting. We clarify statements about the importance of these contributions and conclude that, even for our small gap QW, BIA needs to be taken into account for the proper description of the band structure.
\end{abstract}

\maketitle




\section{Introduction}
\label{sec:k3_intro}

In recent years, interest in developing spin-sensitive devices (spintronics)~\cite{Wolf2000,Gawel2000,DassarmaFabianHu2000,Heinrich2000} has fueled renewed investigations into spin phenomena in semiconductors. The aim is to control not only the spatial degrees of freedom of the electron, but also the spin degree of freedom.  Useful spintronic devices can be devised if such control is achieved. A number of such devices have already been proposed~\cite{DattaDas1990,FiederlingKeimReuscher1999,MorinagaShiiki1999,TingCartoixa2002}. If a full understanding of the operation of spintronic devices is desired, a thorough knowledge of the band structure including all spin details will be needed
. In particular, detailed knowledge of heterostructure bands is required when studying, and perhaps tailoring, time evolution and transport phenomena of spin ensembles.

Most of the standard implementations of the Effective Mass Approximation (EMA) for \iiiv\ heterostructures do not take into account the bulk inversion asymmetry~\cite{Dresselhaus1955} (BIA) present in zincblendes~\cite{MagriZunger2000,TroncKitaev2001}. The BIA lifts Kramers degeneracy and, therefore, a potentially important source of spin splitting in the bands is not included. There have been several proposed modifications to EMA to account for BIA and its associated spin splittings, ranging from 2-band~\cite{EppengaSchuurmans1988:v37} to 14-band~\cite{Rossler1984} and 16-band~\cite{CardonaChristensenFasol1988} models. Also, Zhu and Chang~\cite{ZhuChang1994} started from an 8-band model to generate perturbatively a 2-band Hamiltonian for electrons and a 4-band Hamiltonian for holes, and they performed their calculations of inversion asymmetry effects in that reduced basis set. R\"ossler, Winkler \etal~\cite{WinklerRossler1993,WissingerRosslerWinkler1998} have used an 8-band model solved with a quadrature method that included BIA to study Al$_x$Ga$_{1-x}$As/GaAs quantum wells, where the spin splitting is much smaller than the one shown here. In this work, we study the symmetry properties of the heterostructure Hamiltonian and verify that the 8-band model can account for both the BIA and Rashba~\cite{BychkovRashba1984:physc} effects on the spin splitting.

We present a Finite Difference Method (FDM) implementation of a compact 8-band EMA method capable of reproducing the $D_{2d}$ and $C_{2v}$ symmetry of [001] grown quantum wells and superlattices. We have used this method to study the relative contribution of structural inversion asymmetry (SIA) \vs\ BIA on the spin splitting in the conduction subbands. Our study has allowed us to put into perspective the apparently contradictory statements of Lommer \etal~\cite{LommerMalcherRossler1988}, stating that BIA (SIA) dominates the splitting in large (narrow) gap materials, and Cardona \etal~\cite{CardonaChristensenFasol1988}, listing a proportionality constant for the $k^3$ splitting~\cite{Dresselhaus1955} that becomes roughly larger with smaller bandgap (see Table~\ref{tab:gamma_c}). The FDM allows for arbitrary external fields and the description of tunneling phenomena with only a few changes~\cite{LiuTingMcgill1996}, while maintaining numerical efficiency.

Section~\ref{sec:k3_8band_ema} describes how the 8-band bulk Hamiltonian yields the EMA equations and the method for their solution. In Secs.~\ref{sec:k3_BIA_on_SQWs} and \ref{sec:k3_BIA_on_AQWs} the effects of bulk inversion asymmetry in symmetric and asymmetric quantum wells are explored, taking as example the band structure of AlSb/GaSb/AlSb and AlSb/InAs/GaSb/AlSb quantum wells. Finally, the results are summarized in Sec.~\ref{sec:k3_summary}.

\begin{table}[b]
\centering
\begin{ruledtabular}
\begin{tabular}{ccc}
Material & Band Gap ($eV$) & $\gamma_c$ (\eva3) \\ \hline
GaAs\footnote{Adapted from Ref.~\onlinecite{CardonaChristensenFasol1988}.\label{fn:cardona}} & 1.52 & 25.5 \\
InP$^{ \text{\ref{fn:cardona}} }$ & 1.42 & 8.5 \\
GaSb$^{ \text{\ref{fn:cardona}} }$ & 0.81 & 186.3 \\
InAs\footnote{Ref.~\onlinecite{SilvaLaroccaBassani1994}.} & 0.418 & 130 \\
InSb$^{ \text{\ref{fn:cardona}} }$ & 0.235 & 226.8 \\
\end{tabular}
\end{ruledtabular}
\caption[Band Gap and $\gamma_c$ for selected \iiivs]{Band Gap and $\gamma_c$ for selected \iiivs.}
\label{tab:gamma_c}
\end{table}

\section{EMA Hamiltonian with BIA}
\label{sec:k3_8band_ema}

For the calculation of the heterostructure bands, the Effective Mass Approximation (EMA)~\cite{Bastard1986} based on an 8-band ${\bf k}\cdot{\bf p}$ formalism is used. There are several published 8-band ${\bf k}\cdot{\bf p}$ Hamiltonians~\cite{Kane1957,PidgeonBrown1966,BirPikus}, each including a more or less detailed set of effects. Here we take the Hamiltonian constructed by Trebin {\it et al.}~\cite{TrebinRosslerRanvaud1979} as a starting point. This 
Hamiltonian is constructed by means of an invariant expansion~\cite{Pikus1961:a,Pikus1961:b,BirPikus} and, when applied correctly, guarantees the inclusion of all matrix elements compatible with the $T_d$ symmetry group of the zincblendes up to the desired order in the electron wavevector ${\bf k}$.

For completeness, Appendix~\ref{app:8bandH} reproduces the full 8-band \kp\ Hamiltonian from Ref.~\onlinecite{TrebinRosslerRanvaud1979}. Note that, due to the way that it has been constructed, this Hamiltonian takes into account all the effects of the spin-orbit interaction in the matrix elements up to $k^2$, and in particular the second order $s-p$ coupling via remote $\Gamma_5$ states~\cite{Kane1966} responsible for most of the contribution to the spin splitting in the conduction band. Strain and coupled strain/spin-orbit effects are also properly described by this method. Note that Appendix~\ref{app:8bandH} corrects two typographical errors present in Ref.~\onlinecite{TrebinRosslerRanvaud1979}.

The 8-band Luttinger parameters $\gamma_i$ appearing in Table~\ref{tab:8bandH_elems} are the true Luttinger parameters~\cite{Luttinger1956} $\gamma_{iL}$ with the conduction band contribution subtracted because its effects are treated exactly. They are related by~\cite{PidgeonBrown1966}
\begin{align}
\gamma_1 & = \gamma_{1L} - \frac{1}{3} \frac{E_P}{E_g} \notag \\
\gamma_2 & = \gamma_{2L} - \frac{1}{6} \frac{E_P}{E_g} \label{eq:relKaneLutt} \\
\gamma_3 & = \gamma_{3L} - \frac{1}{6} \frac{E_P}{E_g}, \notag
\end{align}
where $E_g$ is the energy gap of the compound, and $E_P$ has been defined as
\begin{equation}
E_P \equiv \frac{2 m_e P^2}{\hbar^2},
\end{equation}
with $m_e$ being the free electron mass and $P$ the irreducible momentum matrix element.

Table~\ref{tab:parvals} shows the numerical values used in the calculations for the parameters of a number of materials. $B$ and $C$ are parameters describing the BIA effects. As shown below, $B$ is mainly related to the conduction band $k^3$ splitting while $C$ is the coefficient for the linear $k$ splitting in the valence band. No magnetic field effects will be included. For structures grown along the [001] direction, $C_2$, $C_4$ and $C'_5$ will not be needed because there is no shear stress. Finally, the effect of remote bands on the conduction effective mass will be neglected. This amounts to setting the $A'$ parameter to zero.

\begin{table*}[t]
\centering
\begin{minipage}{6.75in}
\begin{ruledtabular}
\begin{tabular}{cccccccccc}
Parameter & InSb & GaSb & AlSb & InAs & GaAs & AlAs \\ \hline
$a$ (\AA)\footnote{Ref.~\onlinecite{Madelung}.\label{fn:madel}} & 6.4794 & 6.096 & 6.136 & 6.058 & 5.653 & 5.66 \\
$E_g$ ($eV$)\footnote{Ref.~\onlinecite{Lawaetz1971}.\label{fn:lawaetz}} & 0.235 & 0.813 & 2.219 & 0.356 & 1.52 & 3.002 \\
$E_v$ ($eV$)\footnote{The valence band offsets are consistent within the systems comprised of (InSb), (GaSb, AlSb, InAs) and (\algaas).} 
   & 0 & 0.56 & 0.11 & 0 & 0 & -0.55 \\
$\Delta_{SO}$ ($eV$)$^{ \text{\ref{fn:lawaetz}} }$ & 0.803 & 0.8 & 0.75 & 0.41 & 0.341 & 0.279 \\
$\gamma_1^{ \text{\ref{fn:lawaetz}} }$ & 2.59 & 2.58 & 1.44 & 2.05 & 2.01 & 1.74 \\
$\gamma_2^{ \text{\ref{fn:lawaetz}} }$ & -0.6 & -0.58 & -0.35 & -0.44 & -0.41 & -0.37 \\
$\gamma_3^{ \text{\ref{fn:lawaetz}} }$ & 0.67 & 0.65 & 0.39 & 0.48 & 0.46 & 0.42 \\
$C$ ($eV\cdot$\AA)\footnote{Ref.~\onlinecite{CardonaChristensenFasol1988}.}
   & -9.32\ee{-3} & 7.00\ee{-4} & 0 & -1.12\ee{-2} & -3.40\ee{-3} & 2.00\ee{-3} \\
$P$ ($eV\cdot$\AA)$^{ \text{\ref{fn:lawaetz}} }$ & 9.35 & 9.21 & 8.41 & 9.17 & 9.86 & 8.94 \\
$B$ ($eV\cdot$\AA$^2$)\footnote{From $\gamma_c$ obtained in Ref.~\onlinecite{PikusMarushchakTitkov1988}.}
   & 10.3 & 49.9 & 0 & 13.7\footnote{From $\gamma_c$ obtained in Ref.~\onlinecite{SilvaLaroccaBassani1994}.} & 30.4
   & 21.3\footnote{From $\gamma_c$ obtained in Ref.~\onlinecite{EppengaSchuurmans1988:v37}.} \\
$C_1$ ($eV$)\footnote{Adapted from the Bir-Pikus deformation potentials $a,b,d$ in Ref.~\onlinecite{Vandewalle1989}.\label{fn:vandewalle}}
   & -6.17 & -6.85 & -6.97 & -5.08 & -7.17 & -5.64 \\
$C_{11}$ (GPa)$^{ \text{\ref{fn:madel}} }$ & 69.18 & 88.34 & 87.69 & 83.29 & 112.6 & 120.2 \\
$C_{12}$ (GPa)$^{ \text{\ref{fn:madel}} }$ & 37.88 & 40.23 & 43.41 & 45.26 & 57.1 & 57 \\
$C_{44}$ (GPa)$^{ \text{\ref{fn:madel}} }$ & 31.32 & 43.22 & 40.76 & 39.59 & 60 & 58.9 
\end{tabular}
\end{ruledtabular}
\end{minipage}
\caption[Parameter values for some materials]{Parameter values for some materials.}
\label{tab:parvals}
\end{table*}

\subsection{Analytical expressions for the bands close to the zone center}
\label{ssec:k3_Energy_expansion}

Starting from the full Hamiltonian shown in Appendix~\ref{app:8bandH}, analytical expressions for the bands near the zone center can be found. These are useful for finding measurable quantities such as effective masses and intraband splittings as a function of the model parameters. These expressions can also be useful when relating the parameters of the model to the parameters used in other families of \kp\ Hamiltonians. The effective masses as a function of the parameters can be found, for example, in Ref.~\onlinecite{EnderleinSipahiScolfaro1998}.

Along [100], the dispersion relation to lowest order in $k$ for the heavy hole (HH) band is
\begin{equation}
E_{HH} \left( {\bf k} \right) = E_v + C k_x,
\label{eq:ehhC}
\end{equation}
while for the light hole (LH) band
\begin{equation}
E_{LH} \left( {\bf k} \right) = E_v - C k_x,
\label{eq:elhC}
\end{equation}
where $E_v$ is the energy at the zone center. Note that each band along [100] is still doubly degenerate as required by the $T_d$ symmetry. In this case $C$ describes a linear splitting between the heavy and light holes. As a consequence of BIA the top of the valence band lies slightly away from $k=0$. This has been experimentally observed~\cite{PidgeonGroves1969,MathurJain1979}.

\begin{figure}[t]
\centering
\epsfig{file=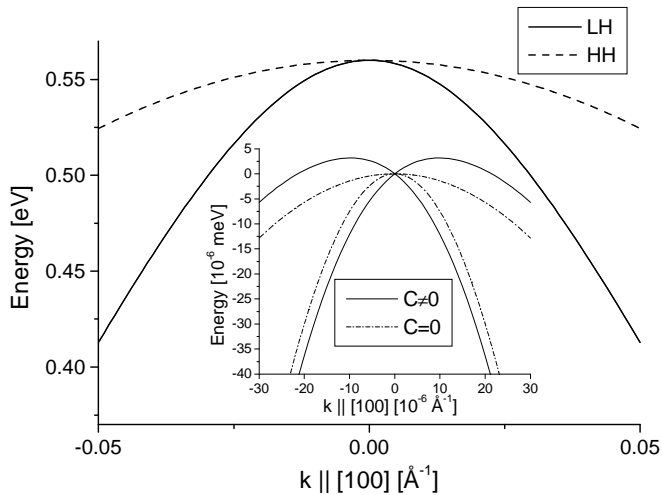, width=\linewidth, clip=}
\caption{HH and LH bands along [100] near the zone center with (BIA) effects included. No difference in the behavior of the bands is appreciable at this scale. The inset shows a blowup of the region very close to the zone center. It is seen that the inclusion of $C \neq 0$ induces the presence of a linear splitting in the vicinity of ${\bf k}=0$. Note the scale of the inset.}
\label{fig:bigC}
\end{figure}

Figure~\ref{fig:bigC} shows the effect of the inclusion of the parameter $C$ in the HH and LH bands along [100] calculated from the numerical diagonalization of \refeq{eq:8bandHAM}. Clearly, the effect is not appreciable at normal scales of {\bf k} and energy. As shown in the inset, we need to look very close to the zone center to observe the effect of $C$. Note the very small value of the energies involved. Away from the zone center the bands with BIA effects recover the usual behavior except for some additional splitting present.

Along the [110] direction, the energy of the electrons in the CB as a function of the wavevector ${\bf k}$ is, up to third order,
\begin{equation}
E_{CB} \left( {\bf k} \right) = E_g + \frac{\hbar^2 k^2}{2m_e m^*_{CB}} \pm \frac{1}{2} \gamma_c k^3,
\label{eq:ECB110}
\end{equation}
with $m^*_{CB}$ being the conduction band effective mass, and the $k^3$ splitting coefficient~\cite{CardonaChristensenFasol1988} $\gamma_c$ given, in terms of the model parameters, by
\begin{equation}
\gamma_c = \frac{P}{3} 
\frac{2B E_g \Delta_{SO} - \sqrt{3} C P \left( E_g + \Delta_{SO} \right)}{E_g^2 \left( E_g + \Delta_{SO} \right)}
\label{eq:gamma_c}
\end{equation}

It is easy to show that the contribution to $\gamma_c$ of the part containing $C$ is only about 4\% for InSb and InAs. That contribution goes down to about 0.3\% for GaAs and AlAs, and it drops to a mere 0.03\% for GaSb. Therefore, it is a good approximation to consider that all the splitting in the conduction band is due to the nonvanishing bulk inversion parameter $B$. Then, if $C$ is neglected, the result of Eppenga \etal~\cite{EppengaSchuurmansColak1987} for the expression of $\gamma_c$ is recovered. Note that, in order to turn off BIA effects, both parameters $B$ and $C$ need to be set to zero.

As in the [100] case, the inclusion of $C$ changes the characteristics of the bands very close to the $\Gamma$ point. Again, it provides them with a small linear component. But, in contrast to the [100] case, here the HH and LH bands are not doubly degenerate. The linear splittings $\Delta_{LH\text{ [110]}}$ and $\Delta_{HH\text{ [110]}}$ turn out to be the same for both HH and LH bands, and are given by
\begin{equation}
\Delta_{HH\text{ [110]}} = \Delta_{LH\text{ [110]}} = \sqrt{3}C k.
\label{eq:HHsplit}
\end{equation}

This result is slightly different from the one indicated in Eq.~(7.5) of Ref.~\onlinecite{CardonaChristensenFasol1988}. A numerical diagonalization of the Hamiltonian has been performed to check the validity of Eq.~(\ref{eq:HHsplit}). The discrepancy arises because the splittings in Ref.~\onlinecite{CardonaChristensenFasol1988} are valid in the region where the quadratic (effective mass) splitting predominates, while the result obtained here is valid in the region where the linear splitting dominates.

The SO band also presents $k^3$ splitting $\Delta_{SO\text{ [110]}}$, proportional to the B parameter only:
\begin{equation}
\Delta_{SO\text{ [110]}} = \frac{2BP}{3 \left( E_g + \Delta_{SO} \right) } k^3.
\end{equation}

Along [111], in the region where the linear splitting dominates, the heavy hole (HH) band has the dispersion relation
\begin{equation}
E_{HH} \left( {\bf k} \right) = E_v \pm \sqrt{2} C k - \frac{\hbar^2 k^2}{2 m_e} \left( \gamma_{1L} - 2\gamma_{3L} \right) + O(k^4),
\end{equation}
while for the light holes
\begin{equation}
E_{LH} \left( {\bf k} \right) = E_v - \frac{\hbar^2 k^2}{2 m_e} \left( \gamma_{1L} + 2\gamma_{3L} \right) + O(k^4).
\end{equation}

The light hole, conduction and split-off bands are degenerate along the [111] direction, as can also be deduced by group theory arguments~\cite{Dresselhaus1955}. The linear splitting obtained for the heavy holes agrees with the result of Cardona \etal~\cite{CardonaChristensenFasol1988}.

\subsection{Bulk bands and spin behavior}

\begin{figure}[t]
\centering
\epsfig{file=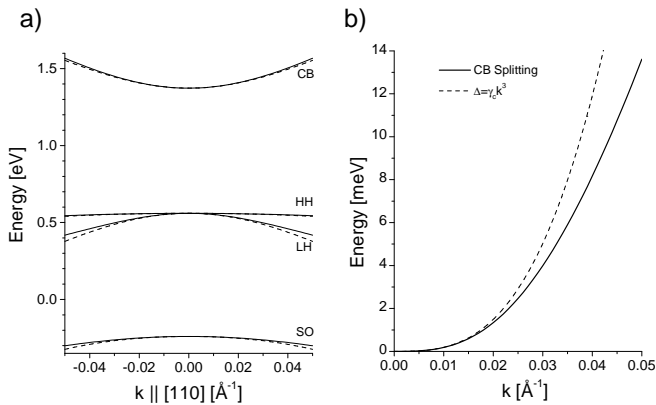, width=\linewidth, clip=}
\caption[Band structure for GaSb near $\Gamma$ along [110\rsqbkt.]{Band structure for GaSb near $\Gamma$ along [110] and spin splitting of the conduction band. The dashed line of plot b) corresponds to the function $\text{Splitting}=\gamma_c k^3$, with $\gamma_c = 186.3$ \eva3.}
\label{fig:bands_gasb110}
\end{figure}

An example of bulk bands computed using the Hamiltonian with the full symmetry can be seen in Fig.~\ref{fig:bands_gasb110}.a), which shows the band structure of GaSb along the [110] direction. For this direction, all bands are spin split except at the $\Gamma$ point. Plot b) shows the energy splitting of the CB states as a function of $k$. Close to the zone center, the splitting follows the behavior described in \refeq{eq:ECB110}, with $\gamma_c$ taken to be the experimental value. In this case, it is seen that the usual 2-band Hamiltonian model [see \refeq{eq:Ham_k3} below] can describe the $k^3$ splitting with good accuracy up to $~2.5\%$ of the zone boundary.

This $k^3$ splitting~\cite{Dresselhaus1955} can also be predicted by the methods described in Refs.~\onlinecite{Pikus1961:a,Pikus1961:b}. Going up to order 3 in the combinations of components of ${\bf k}$ and constructing an invariant 2-band Hamiltonian for the conduction band, it is found that the Hamiltonian will include the following term breaking the spin degeneracy~\cite{SilvaLaroccaBassani1994}:
\begin{equation}
H_{k^3} = \gamma_c \left[ \sigma_x k_x \left( k_y^2 - k_z^2  \right) + \text{cyclic permutations} \right].
\label{eq:Ham_k3}
\end{equation}

\begin{figure}[t]
\centering
\epsfig{file=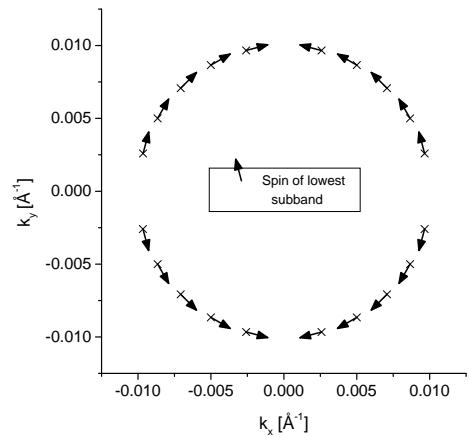, width=0.7\linewidth, clip=}
\caption[Spin directions for a conduction subband of GaSb.]{Direction for the spin of the spin-split states of the lowest conduction subband of GaSb. This plots sweeps a circular path in ${\bf k}$ space with $k_z=0$. No spin direction is specified for the $\langle 100\rangle$ family because the states are spin degenerate.}
\label{fig:spinwheel_bulk_GaSb}
\end{figure}

The two band Hamiltonian in \refeq{eq:Ham_k3} also predicts the direction where the split spins will be pointing. For example, it is easily seen from the previous equation that if ${\bf k}=(1,\delta,0)$, with $\delta$ a positive infinitesimal, the spin will point along the $\pm y$ direction. Similarly, symmetry requires the spin to point along $(-1,1,0)$ or $(1,-1,0)$ for ${\bf k}$ along the [110] direction. This is indeed obtained in the numerical diagonalization of the Hamiltonian including the BIA effects, as seen in Fig.~\ref{fig:spinwheel_bulk_GaSb}. That figure shows a circular sweep in ${\bf k}$ space with $k_z=0$. The arrows represent the direction towards which the spin of the lowest conduction subband states is pointing. The horizontal axis represents $k_x$, while the vertical axis indicates the $k_y$ component of the state. The states belonging to the $\langle 100\rangle$ directions are spin degenerate; therefore, no spin direction is given for them in Fig.~\ref{fig:spinwheel_bulk_GaSb}.

\subsection{Interface conditions and Hermiticity in the FDM}

In Appendix~\ref{sec:k3_FDM} we describe the Finite Difference Method (FDM) we use to find the heterostructure bands and eigenstates. This method transforms the EMA set of coupled ordinary differential equations into the eigenvalue problem shown in \refeq{eq:eigensystem}.

The hermiticity of the discretized Hamiltonian operator in \refeq{eq:eigensystem} is ensured if ${\bf H}^\dagger_{i,i+1} = {\bf H}_{i+1,i}$. Since the ${\bf H}^{(j)}$'s are themselves Hermitian, an inspection of \refeqs{eq:Hii}{eq:Himi} shows that this is indeed the case. It is also clearly seen that the introduction of quantum well or superlattice boundary conditions (BCs) doesn't affect the Hermiticity of the Hamiltonian.

\begin{figure}[t]
\centering
\epsfig{file=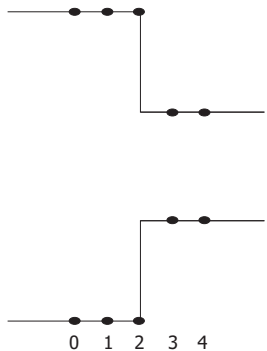, width=0.4\linewidth, clip=}
\caption[Mesh used in the study of interface conditions.]{Mesh used in the study of interface conditions.}
\label{fig:interface}
\end{figure}

There exist in the literature several proposals on what are the correct quantities to match at the interface between two materials~\cite{EppengaSchuurmansColak1987,YamanakaKamadaYoshikuni1994,Foreman1997,ChuangChang1997}. Most of them require the continuity of the envelope function and a quantity that has the general form:
\begin{equation}
\left[ {\bf A} \partial_z + {\bf B} \right] {\bf F}
\label{eq:int_cond}
\end{equation}
where the matrices ${\bf A}$ and ${\bf B}$ take different values depending on the author. Using the finite difference formulae \refeqs{eq:fd_H2}{eq:fd_H1}, the continuity of \refeq{eq:int_cond} can be written in a form similar to \refeq{eq:fd_eqs}:
\begin{equation}
{\bf H}_{2,1} {\bf F}_{1} + {\bf H}_{2,2} {\bf F}_{2} + {\bf H}_{2,3} {\bf F}_{3} = 0
\label{eq:interface}
\end{equation}
where the subindexes $i,j$ are referred to the mesh points in Fig.~\ref{fig:interface} and the ${\bf H}_{i,j}$ are given by
\begin{align}
{\bf H}_{2,1} &= + i \frac{{\bf A}^L - {\bf A}^R}{4\Delta z} \notag \\
{\bf H}_{2,2} &= {\bf B}^L - {\bf B}^R \notag \\
{\bf H}_{2,3} &= - i \frac{{\bf A}^L - {\bf A}^R}{4\Delta z} \label{eq:Hinterface},
\end{align}
with $L$ ($R$) meaning the material at the left (right) of the interface. Isolating ${\bf F}_{2}$ from \refeq{eq:interface} and plugging it into the corresponding equations for the ${\bf F}_{i}$'s, one obtains
\begin{gather}
{\bf H}_{1,0} {\bf F}_{0} + \left( {\bf H}_{1,1} - {\bf H}_{1,2} {\bf H}^{-1}_{2,2} {\bf H}_{2,1} \right) {\bf F}_{1} - \notag \\
{\bf H}_{1,2} {\bf H}^{-1}_{2,2} {\bf H}_{2,3} {\bf F}_{3} = E \: {\bf F}_{1} \\
- {\bf H}_{3,2} {\bf H}^{-1}_{2,2} {\bf H}_{2,1} {\bf F}_{1} + \left( {\bf H}_{3,3} -
 {\bf H}_{3,2} {\bf H}^{-1}_{2,2} {\bf H}_{2,3} \right) {\bf F}_{3} + \notag \\
{\bf H}_{3,4} {\bf F}_{4} = E \: {\bf F}_{4}
\end{gather}

Now, in order to preserve the Hermiticity of the discretized Hamiltonian, one should have
\begin{equation}
{\bf H}^\dagger_{2,3} = {\bf H}_{3,2} \qquad {\bf H}^\dagger_{2,1} = {\bf H}_{1,2}.
\label{eq:req_Herm}
\end{equation}

However, looking at Eqs.~(\ref{eq:Hinterface}) and \refeqs{eq:Hii}{eq:Himi}, one sees that the requirements in \refeq{eq:req_Herm} are not satisfied by the discretized version of the interface conditions. From this it must be concluded that the enforcement of interface conditions of the form (\ref{eq:int_cond}) is not possible if the Hermiticity of the Hamiltonian is to be preserved, and that \refeqs{eq:Hii}{eq:Himi} must be used instead for all mesh points.

\subsection{Symmetry group of the discretized Hamiltonian}
\label{sec:k3_sym_disc_Ham}

\begin{table}[b]
\centering
\begin{ruledtabular}
\begin{tabular}{cccc}
Point/Line & \parbox{2cm}{Point Group Symmetry} & \parbox{1.5cm}{Spin Splitting} & Spin directions \\
\hline
$\Gamma$ & $D_{2d}$ & No & \\
$\Delta$ [100] & $C_2$ & Yes & [100],[$\bar{1}$00] \\
$\Sigma$ [110] & $C_s$ & Yes & [$\bar{1}$10],[1$\bar{1}$0] \\
Other points & $C_1$ & Yes & Undetermined
\end{tabular}
\end{ruledtabular}
\caption[Symmetry requirements on degeneracy and spin splitting for a $D_{2d}$ structure.]{Symmetry requirements on spin splitting and directions for points in the $k_x-k_y$ plane in a $D_{2d}$ structure.}
\label{tab:sym_D2d}
\end{table}

For the rest of this article, we will refer to ``symmetric'' and ``asymmetric'' heterostructures, mainly quantum wells (QWs). Here, ``symmetric'' will be taken to mean that the sequence of materials and their respective thicknesses are left unchanged under the inversion operation (\ie, they are macroscopically symmetric). Thus, an AlSb/InAs/AlSb QW is called symmetric, while an AlSb/InAs/GaSb/AlSb QW is called asymmetric. This definition is made in order to avoid confusion with the microscopic symmetry --that is, the symmetry group-- of the QW. All asymmetric [001] heterostructures made from zincblendes are described by the $C_{2v}$ point group. On the other hand, symmetric [001] QWs can belong to either the $C_{2v}$ or the $D_{2d}$ symmetry groups depending on an interplay of characteristics such as the parity of the number of monolayers, the existence of a common atom in the constituents~\cite{WangWeiMattila1999} or the presence of a supplementary half-layer~\cite{Vogl2001}.

The symmetry group of the discretized Hamiltonian in \refeq{eq:eigensystem} can be found by verifying that all operations $g$ of the $T_d$ point group satisfy the relationship~\cite{BirPikus}
\begin{equation}
D^{-1}(g) {\bf H} ({\bf k}_\parallel) D(g) = {\bf H} ( g^{-1} {\bf k}_\parallel ) ,
\end{equation}
where $D(g)$ is the representation of the $g$ operator in the basis of the discretized Hamiltonian ${\bf H} ({\bf k}_\parallel)$, and ${\bf k}_\parallel=(k_x,k_y)$. This tedious procedure can be done with the help of computer software, such as {\em Mathematica}~\cite{Mathematica}, which automates algebraic manipulations. It is seen that the EMA Hamiltonian corresponding to symmetric structures transforms according to $D_{2d}$, while for asymmetric structures it transforms according to $C_{2v}$. This is in contrast to the majority of EMA implementations, which lack the inclusion of bulk inversion asymmetry effects and reproduce an approximate $D_{4h}$ symmetry~\cite{TroncKitaev2001} for symmetric structures.

\begin{table}[b]
\centering
\begin{ruledtabular}
\begin{tabular}{cccc}
Point/Line & \parbox{2cm}{Point Group Symmetry} & \parbox{1.5cm}{Spin Splitting} & Spin directions \\
\hline
$\Gamma$ & $C_{2v}$ & No & \\
$\Delta$ [100] & $C_1$ & Yes & Undetermined \\
$\Sigma$ [110] & $C_s$ & Yes & [$\bar{1}$10],[1$\bar{1}$0] \\
Other points & $C_1$ & Yes & Undetermined
\end{tabular}
\end{ruledtabular}
\caption[Symmetry requirements on degeneracy and spin splitting for a $C_{2v}$ structure.]{As in Table~\ref{tab:sym_D2d}, for a $C_{2v}$ structure.}
\label{tab:sym_C2v}
\end{table}

\begin{figure}[t]
\centering
\epsfig{file=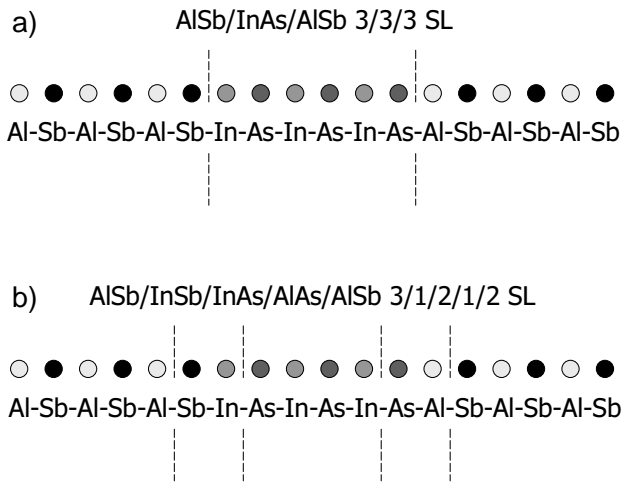, width=0.95\linewidth, clip=}
\caption[Layer arrangement for a no-common-atom quantum well.]{Layer arrangements for a no-common-atom quantum well. An arrangement as in a) in the EMA would yield $D_{2d}$ symmetry. The alternative arrangement b) yields the correct $C_{2v}$ symmetry of the heterostructure.}
\label{fig:ncaqw}
\end{figure}

Tables~\ref{tab:sym_D2d} and \ref{tab:sym_C2v} show the requirements that the underlying symmetry of the atom arrangement imposes on the spin degeneracy of the energy levels and the direction where the spins are pointing in case the levels are not degenerate. Note that, for a structure with $D_{2v}$ symmetry (\ie, BIA effects only) the bands along [100] must be split~\cite{EppengaSchuurmans1988:v37} even though there is no splitting along [100] for bulk zincblendes.

Although the \kp\ method is not designed taking into account the interface characteristics of no-common-atom (NCA) heterostructures, it is possible in some situations to modify the simulated structure to obtain at least the right symmetry effects. Figure~\ref{fig:ncaqw}.a) shows a NCA quantum well. The way that the boundaries of the layers are set up, the well would be symmetric and, therefore, the Hamiltonian would have $D_{2d}$ symmetry instead of the $C_{2v}$ corresponding to the asymmetric interface bonds. However, the material boundaries in the \kp\ method are arbitrary to half a monolayer. As seen in Fig.~\ref{fig:ncaqw}.b), a simple rearrangement of the material boundaries reproduces the asymmetry in the bonds and allows us to take into account, at least qualitatively, the effects of the lower symmetry.

The only case that cannot be modeled through these rearrangements is when a common atom structure, such as an AlAs/GaAs/AlAs QW, has $C_{2v}$ symmetry due to the well having an odd number of monolayers. In that case, even though the species participating at the bond at the interface are the same, there is an asymmetry in the bond orientation, which the present EMA method cannot take into account.

With the control that the proposed model allows over the symmetry of the Hamiltonian, it is possible to use the EMA in studies of the spin splitting appearing in heterostructures. Thus one can delineate the role of bulk inversion asymmetry \vs\ structure inversion asymmetry, layer asymmetry \vs\ interface asymmetry, etc. This model also provides a straightforward and easy to implement tool for studying reduced symmetry effects, such as the presence of optical anisotropy~\cite{KrebsVoisin1996,MagriOssicini2001}, and the mixing of heavy hole and light hole states at the zone center~\cite{MagriZunger2000,IvchenkoKaminskiRossler1996} in QWs and superlattices. Some of these works have followed the alternative approach of introducing interface parameters to describe the lowering of symmetry respect to the standard EMA formalism~\cite{IvchenkoKaminskiRossler1996,KrebsVoisin1996,VervoortFerreiraVoisin1997}. In these cases interface asymmetry effects are well described, but BIA is not accounted for.

\section{Bulk Inversion Asymmetry Effects on symmetric Quantum Wells}
\label{sec:k3_BIA_on_SQWs}

In this section, the methods described in Sec.~\ref{sec:k3_8band_ema} will be used to calculate the electronic properties of a symmetric quantum well. In particular, focus will fall on AlSb/GaSb/AlSb quantum wells. However, some of the results derived are a consequence of the underlying symmetry of the structure rather than the constituents themselves. Therefore, these particular results will illustrate general considerations.

\subsection{SQWs without BIA terms}
\label{sec:k3_symQW_noBIA}


\begin{figure}[t]
\centering
\epsfig{file=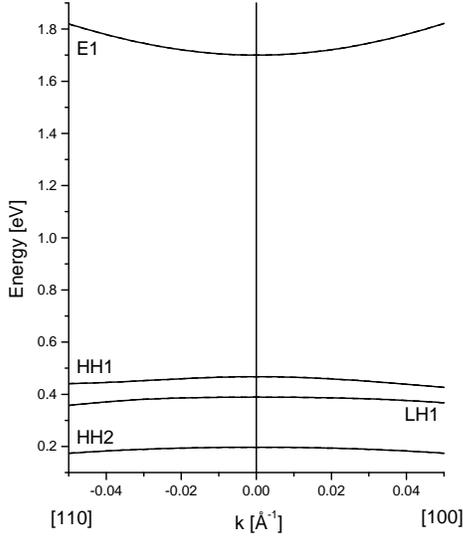, width=0.7\linewidth, clip=}
\caption[Bands for a symmetric quantum well w/o BIA terms.]{Bands along [100] and [110] for an AlSb/GaSb/AlSb SQW 8 monolayers thick without BIA terms.}
\label{fig:bs_symQW_noBIA}
\end{figure}

Figure~\ref{fig:bs_symQW_noBIA} shows the band structures along the [100] and the [110] directions of a common atom AlSb/GaSb/AlSb symmetric quantum well (SQW) grown along the [001] direction and with a well thickness of 8 monolayers (24.4 \AA). Since no inversion asymmetry affects are included, the bands show Kramers degeneracy throughout the Brillouin zone and the quantization axes of the spins are not unambiguously defined.

The labels E1, HH1, LH1 and HH2 shown in the plots correspond to the first electron, first heavy hole, first light hole and second heavy hole states in the QW respectively. They refer to the main bulk state contribution at ${\bf k}=0$. For a well without BIA terms and in the zone center, the heavy holes decouple from the rest of the bands, and the HH$n$ states have only bulk heavy hole components. This is in contrast to the E$n$ (LH$n$) bands, which have small bulk light hole (electron) and split off contributions even at the zone center due to the loss of translational symmetry along [001] caused by the well potential.

\subsection{SQWs with BIA terms}
\label{sec:k3_symQW_BIA}

\begin{figure}[t]
\centering
\epsfig{file=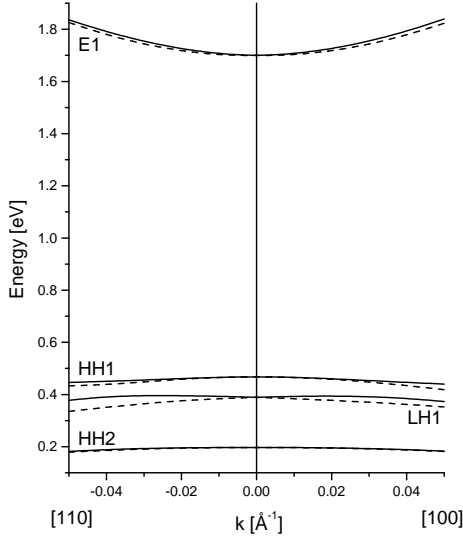, width=0.7\linewidth, clip=}
\caption[Bands for a symmetric quantum well with BIA terms.]{Same as Fig.~\ref{fig:bs_symQW_noBIA}, but with BIA terms.}
\label{fig:bs_symQW_BIA}
\end{figure}

\begin{figure}[t]
\centering
\epsfig{file=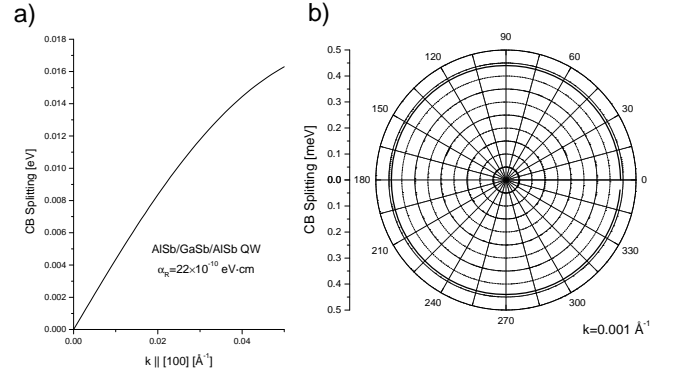, width=0.98\linewidth, clip=}
\caption[Linear and angular spin splitting dependence for a SQW.]{a) Spin splitting dependence for an AlSb/GaSb/AlSb QW along the [100] line. The calculated proportionality constant is $\alpha_{BIA}=22 \times 10^{-10}$ eV$\cdot$cm. b) Splitting dependence along a circle in the $k_x-k_y$ plane, with $k=0.001$ \AA.}
\label{fig:sp_split_symQW}
\end{figure}

Figure~\ref{fig:bs_symQW_BIA} shows the same band structures as in Fig.~\ref{fig:bs_symQW_noBIA}, but with BIA terms included. As predicted by group theory (cf. Table~\ref{tab:sym_D2d}), the bands are split along both directions except at the zone center. This is the major difference with most of the EMA models in the literature.

With the inclusion of BIA terms, the heavy hole states couple with the light holes by means of remote states through a perturbative mixed spin orbit and \kp\ interaction parametrized by $C$~\cite{CardonaChristensenFasol1988}. Thus, the HH$n$ states lose their pure bulk heavy hole character and, in particular, the HH{\em even} (HH{\em odd}) mix with the LH{\em odd} (LH{\em even})~\cite{MagriZunger2000}. However, looking at the wavefunction for the HH1 state of the structure in Fig.~\ref{fig:bs_symQW_BIA}, it can be seen that the contribution from bulk components other than the HH to the probability density is about 8 orders of magnitude less than the heavy hole contribution.

Figure~\ref{fig:sp_split_symQW} shows the linear behavior and the isotropy of the conduction subband spin splitting close to the zone center. Plot a) shows the dependence of the spin splitting in the CB along the [100] line. It is seen that the splitting is linear close to the $\Gamma$ point, with a proportionality coefficient of $\alpha_{BIA}=22 \times 10^{-10}$ eV$\cdot$cm, where the splitting is
\begin{equation}
\Delta_{BIA} = 2 \alpha_{BIA} k.
\end{equation}

Although there is an isotropic linear spin splitting, this must not be confused with the Rashba splitting~\cite{BychkovRashba1984:physc}. Bychkov and Rashba introduced a splitting coefficient $\alpha_R$ in the context of asymmetric quantum wells. The splitting studied there is derived from a model Hamiltonian that describes only structural inversion asymmetry (SIA) effects, but not bulk inversion asymmetry. As a consequence, the spin directions that they predict don't apply to the SQW situation (cf. Sec.~\ref{sec:k3_BIA_on_AQWs}).

The computed $\alpha_{BIA}$ for this structure is about half of some of the highest predicted Rashba coefficients for asymmetric structures~\cite{CartoixaTingDaniel2001,CartoixaTingMcgill2002b} where only SIA contributions are taken into account. This shows that BIA effects need to be carefully studied before neglecting them in a calculation.

A model Hamiltonian for spins in the conduction subbands of SQWs in the same spirit as the Rashba Hamiltonian~\cite{BychkovRashba1984:physc} was obtained by de Andrada e Silva~\cite{Silva1992}
\begin{equation}
H = \alpha_{BIA} \left( \sigma_x k_x - \sigma_y k_y \right),
\label{eq:HR_sym}
\end{equation}
where $\sigma$ are the Pauli matrices, $\alpha_{BIA} = -\melem{F_s;k_x,k_y}{\gamma_c \partial_z^2}{F_s;k_x,k_y}$ and $\ket{F_s;k_x,k_y}$ is the envelope function corresponding to the electron traveling in the plane with wavevector $k_x,k_y$.

\begin{figure}[t]
\centering
\epsfig{file=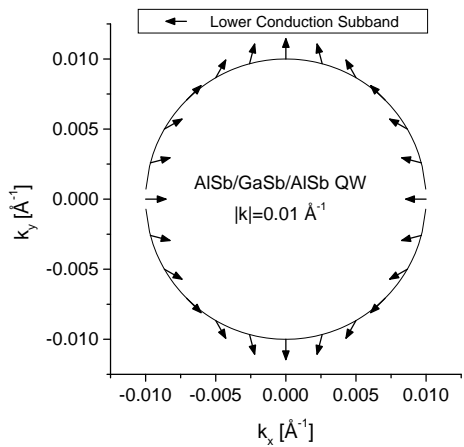, width=0.7\linewidth, clip=}
\caption[Spin directions for the lower CB of an AlSb/GaSb/AlSb QW.]{Spin directions for the lower conduction subband of an AlSb/GaSb/AlSb SQW. The thickness of the well is 8\ML\ (24.4 \AA). The spins are plot at 15\deg\ intervals, and correspond to states lying on a circle in the $k_x-k_y$ plane with $k=0.01$ \AA.}
\label{fig:sp_wheel_symQW}
\end{figure}

The behavior of the spins when BIA terms are included is very interesting. Figure~\ref{fig:sp_wheel_symQW} shows the spin directions of the eigenstates of the lowest conduction subband from our calculations. The spin directions are shown for a circular sweep in the $k_x-k_y$ plane keeping $k=0.01$ \AA. The directions of the spins agree with what would be predicted from \refeq{eq:HR_sym}. The spins at a given point in the $k_x-k_y$ plane point in opposite directions for the two subbands. Note that in a given subband, although the $x$ and $y$ axes are equivalent, in one of the axes the spin points outward while in the other it points inward. The explanation lies in the way that the $x$ and $y$ axes are connected and in the fact that spinors don't change sign under inversion. For a QW with $D_{2d}$ symmetry, the $x$ and $y$ axes are equivalent through a reflection by the plane containing the [110] and [001] directions. The reflection by this plane can be thought of as a rotation of 180$^\circ$ along the [$\bar{1}$10] direction followed by an inversion. Starting with a state with $k_y$ component only $\ket{k_y,\uparrow_{\hat{y}}}$ (spin pointing outward), the rotation will send it to $\ket{-k_x,\uparrow_{-\hat{x}}}$ (still outward). Then, the inversion will flip ${\bf k}$, but not the spin, sending the state to $\ket{k_x,\uparrow_{-\hat{x}}}$ (spin pointing inward).

\section{Bulk Inversion Asymmetry Effects on asymmetric Quantum Wells}
\label{sec:k3_BIA_on_AQWs}


In this section the structure under study will be an AlSb/InAs/GaSb/AlSb asymmetric quantum well (AQW) grown along the [001] direction compliant with a GaSb substrate. The thickness of the InAs and GaSb layers is 8 monolayers (ML) each, with a monolayer thickness of 3.048 \AA.

\subsection{AQWs without BIA terms}
\label{sec:k3_asymQW_noBIA}

\begin{figure}[t]
\centering
\epsfig{file=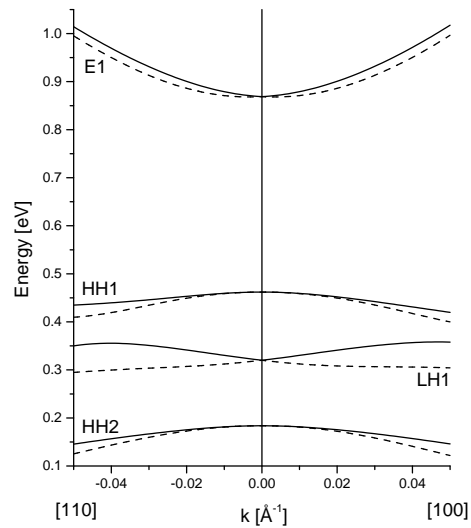, width=0.7\linewidth, clip=}
\caption[Bands for an asymmetric quantum well w/o BIA terms.]{Bands without BIA effects for an AlSb/InAs/GaSb/AlSb AQW grown along the [001] direction compliant with a GaSb substrate. The bands are along [100] and [110]. The thickness of the InAs and GaSb layers is 8 ML each.}
\label{fig:bs_asymQW_noBIA}
\end{figure}

\begin{figure}[t]
\centering
\epsfig{file=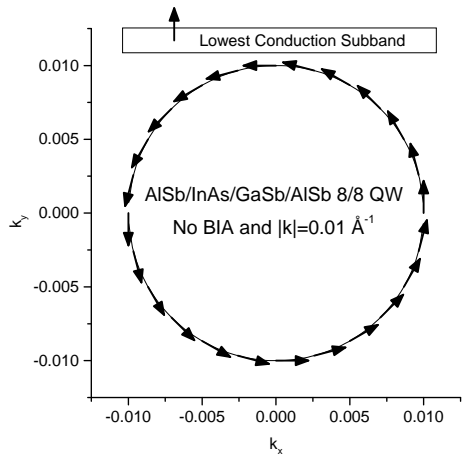, width=0.7\linewidth, clip=}
\caption[Spin directions for a 2DEG in an AlSb/InAs/GaSb/AlSb quantum well]{Spin directions for the lower conduction subband of an AlSb/InAs/GaSb/AlSb 8/8 AQW {\em without} BIA effects included. The spins are plot at 15\deg\ intervals, and correspond to the lowest conduction subband states lying on a circle in the $k_x-k_y$ plane with $k=0.01$ \AA.}
\label{fig:spinwheel}
\end{figure}

Figure~\ref{fig:bs_asymQW_noBIA} shows the band structure along the [100] and the [110] directions of the AQW {\em without} the inclusion of BIA terms. The structural inversion asymmetry (SIA) reduces the symmetry group from $D_{4h}$ to $C_{2v}$, and a spin splitting appears between the conduction subbands. The splitting due to SIA effects is usually modeled using a Hamiltonian first introduced by Bychkov and Rashba~\cite{BychkovRashba1984:physc}:
\begin{equation}
H_R = \alpha_R \left( \bm{\sigma} \times {\bf k} \right) \cdot \bm{\nu},
\label{eq:HR_asym}
\end{equation}
where $\alpha_R$ is the Rashba coefficient, $\bm{\sigma}$ is a vector composed of the Pauli matrices and $\bm{\nu}$ is the axis of symmetry of the structure. This Hamiltonian is valid for describing the SIA contributions close to the zone center. It predicts a linear and isotropic splitting
\begin{equation}
\Delta_R = 2 \alpha_R k,
\end{equation}
where $k$ is the magnitude of the electron wavevector. It also predicts that the spins will point tangentially to the circles of constant $k$ in the $k_x-k_y$ plane, which is verified by numerical calculations (see Fig.~\ref{fig:spinwheel}).

\subsection{AQWs with BIA terms}
\label{sec:k3_asymQW_BIA}

\begin{figure}[t]
\centering
\epsfig{file=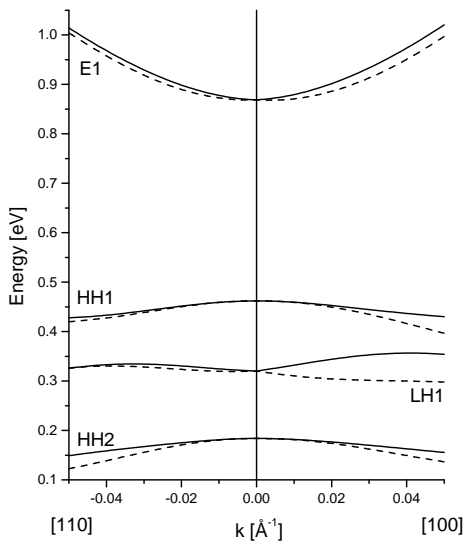, width=0.7\linewidth, clip=}
\caption[Bands for an asymmetric quantum well with BIA terms.]{Bands along [100] and [110] for an AlSb/InAs/GaSb/AlSb AQW with BIA effects.}
\label{fig:bs_asymQW_BIA}
\end{figure}

The band structure for the asymmetric quantum well (AQW) under study {\em with} BIA effects is shown in Fig.~\ref{fig:bs_asymQW_BIA}. The effects of inversion asymmetry are highly anisotropic in bulk~\cite{Dresselhaus1955}, and this reflects on the directional dependence of the bands. Comparing with Fig.~\ref{fig:bs_asymQW_noBIA}, it is seen that the BIA effects are necessary to obtain accurate bands in the [110] direction. In particular, note that the LH1 splitting along [110] is much smaller when BIA effects are included. However, we should point out that the LH1 splitting, with BIA, along [$\bar{1}$10] becomes much larger due to the inequivalency of the [110] and [$\bar{1}$10] under the $C_{2v}$ point group.

The inclusion of the Hamiltonian (\ref{eq:HR_sym}) keeps the analysis of the combined BIA and SIA effects quite simple. For a [001] structure, the SIA and BIA contributions to the splitting can be described by a Hamiltonian $H_{\text{IA}}$ made from the addition of \refeq{eq:HR_sym} and \refeq{eq:HR_asym}:
\begin{multline}
H_{\text{IA}} = \alpha_{\text{BIA}} \left( \sigma_x k_x - \sigma_y k_y \right) + \alpha_R \left( \sigma_x k_y - \sigma_y k_x \right) = \\
\sigma_x \left( \alpha_R k_y + \alpha_{\text{BIA}} k_x \right) - \sigma_y \left( \alpha_{\text{BIA}} k_y + \alpha_R k_x \right),
\label{eq:HR_IA}
\end{multline}
where $\alpha_{\text{BIA}}$ ($\alpha_R$) is the coefficient describing BIA (SIA) effects. From here, making an analogy with the Zeeman splitting, it is easy to find that the splitting in the conduction band (CB) close to the zone center will be
\begin{equation}
\Delta_{\text{IA}} = 2 k \sqrt{\alpha_R^2 + 2 \alpha_R \alpha_{\text{BIA}} \sin 2\theta + \alpha_{\text{BIA}}^2},
\label{eq:split_IA}
\end{equation}
where $\theta$ is the in-plane polar angle.

\begin{figure}[t]
\centering
\epsfig{file=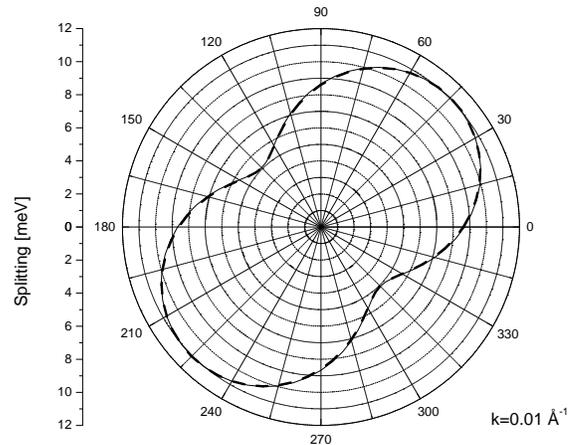, width=0.85\linewidth, clip=}
\caption[Angular dependence of the spin splitting for an AQW.]{Angular dependence of the spin splitting for an AlSb/InAs/GaSb/AlSb AQW. The solid line is the 8-band model numerical result. The dashed line is a fit using \refeq{eq:split_IA} with $\alpha_R=40.3\times 10^{-10}$ eV$\cdot$cm and $\alpha_{\text{BIA}}=15.0\times 10^{-10}$ eV$\cdot$cm.}
\label{fig:split_polar_asymQW}
\end{figure}

A full 8-band numerical calculation of the splitting along a circle in the $k_x-k_y$ plane and the 2-band prediction from expression (\ref{eq:split_IA}) are shown in Fig.~\ref{fig:split_polar_asymQW}. The values from the analytic expression show very good agreement with the numerical results from our 8-band implementation. The numerical results are fitted with $\alpha_{\text{SIA}}=40.3\times 10^{-10}$ eV$\cdot$cm and $\alpha_{\text{BIA}}=15.0\times 10^{-10}$ eV$\cdot$cm. This way the BIA effects are quantified, and it must be concluded that they need to be taken into account for an accurate description of the bands. This is clearly so in the [110] direction, where the contributions are added linearly, but it is also true in a lesser degree in the [100] direction, where the contributions are added quadratically. This also has a clear effect on the theory of extraction of the Rashba coefficient from Shubnikov-de Haas measurements in two-dimensional electron gases (2DEGs), where BIA effects have been usually neglected~\cite{LuoMunekataFang1988,LuoMunekataFang1990,SchapersEngelsLange1998}.

For a quantum well where the 2-band model is valid, the BIA splitting coefficient for the CB can be estimated with~\cite{EppengaSchuurmans1988:v37}
\begin{equation}
\alpha_{\text{BIA}} \approx \frac{\gamma_{cW}}{L_W^2},
\end{equation}
where $\gamma_{cW}$ and $L_W^2$ are the $k^3$ splitting coefficient of the CB and the thickness of the layer where the electrons are confined respectively. This estimate will become more accurate as the well becomes thicker. From this expression and \refeq{eq:gamma_c} it is readily seen that BIA effects will be considerable when the material in the well layer has a low bandgap and high spin-orbit interaction, such as InAs, GaSb and InSb (cf. Table~\ref{tab:gamma_c}) and in a narrow well. So, it has been seen that, a priori, it is not possible to consider only SIA effects for an asymmetric structure even if the constituents are low bandgap materials. The statement of Lommer \etal~\cite{LommerMalcherRossler1988} about BIA (SIA) effects dominating in large (narrow) gap systems must be understood as referring to the dependence on the gap of the prefactors $\gamma_c$ and the part of $\alpha_R$ not proportional to the expectation value of the electric field. But other factors such as the well width or an applied bias~\cite{MajewskiVogl2002} can change the relative contributions.

\begin{figure}[t]
\centering
\epsfig{file=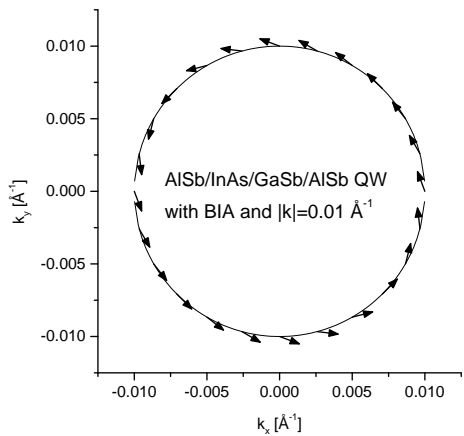, width=0.7\linewidth, clip=}
\caption[Spin directions corresponding to the lower CB of an AlSb/InAs/GaSb/AlSb QW.]{Spin directions for the lower conduction subband of an AlSb/InAs/GaSb/AlSb AQW {\em with} BIA effects. The spins are plot at 15\deg\ intervals, and correspond to the lowest conduction subband states lying on a circle in the $k_x-k_y$ plane with $k=0.01$ \AA.}
\label{fig:sp_wheel_asymQW_BIA}
\end{figure}

Finally, the electron spins are also affected by the inclusion of BIA terms. In Fig.~\ref{fig:sp_wheel_asymQW_BIA} the spins of the lowest conduction subband are shown for states lying on a circle in the $k_x-k_y$ plane of radius $k=0.01$ \AA. The direction of the spins has changed respect to the case without BIA terms (see Fig.~\ref{fig:spinwheel}). As it can be deduced from \refeq{eq:HR_IA}, it corresponds to the vector sum of the spins in Fig.~\ref{fig:spinwheel} and the spins in Fig.~\ref{fig:sp_wheel_asymQW_BIA}, each one weighted by their corresponding splitting coefficients $\alpha$'s.

\begin{table*}
\centering
\setlength{\extrarowheight}{4pt}
\begin{tabular}{c@{=}l}
\hline \hline
$H^{cc}$ & $E_v + E_g + \frac{\hbar^{2^{\parbox[t][2mm][t]{1pt}{} }} k^2}{2 m} + A' k^2 - g_s \frac{e \hbar}{4 m c} \bm{\sigma} \cdot {\bf H} + C_1 \tr \overline{\overline{\epsilon}}$ \\
$H^{vv}$ & $H^{vv}_k + H^{vv}_\epsilon + H^{vv}_{kl} + H^{vv}_{\epsilon k}$ \\
$H^{vv}_k$ & $-\frac{\hbar^2}{m} \left\{ \frac{1}{2} \gamma_1 k^2 - \gamma_2 \left[ \left( J_x^2 - \frac{1}{3} J^2 \right) k_x^2 + cp \right]
-2 \gamma_3 \left[ \left\{ J_x J_y \right\} \left\{ k_x k_y \right\} + cp \right] \right\} -
\frac{e \hbar}{m c} \left\{ \left( \kappa J_x + q J_x^3 \right) H_x + cp \right\}$ \\
$H^{vv}_\epsilon$ & $D_d \tr \overline{\overline{\epsilon}} + \frac{2}{3} D_u \left[ \left( J_x^2 - \frac{1}{3} J^2 \right) \epsilon_{xx} +
cp \right] + \frac{2}{3} D_u' \left[ 2 \left\{ J_x J_y \right\} \epsilon_{xy} + cp \right] $ \\
$H^{vv}_{kl}$ & $ \frac{2}{\sqrt{3}} C \left[ \left\{ J_x \left( J_y^2 - J_z^2 \right) \right\} k_x + cp \right] $ \\
$H^{vv}_{\epsilon k}$ & $ \left[ C_4 \left( \epsilon_{yy} - \epsilon_{zz} \right) k_x + C_5' \left( \epsilon_{xy} k_y - \epsilon_{xz} k_z \right)
\right] J_x + cp $ \\
$H^{ss}$ & $ - \Delta_{so} + \frac{\hbar^2}{2m} \gamma_1 k^2 - 2 \kappa \frac{e \hbar}{2 m c} \bm{\sigma} \cdot {\bf H} + D_d \tr \overline{\overline{\epsilon}}$ \\
$H^{cv}$ & $ \sqrt{3} \left[ P \left( k_x T_x + cp \right) + i B \left( T_x \left\{ k_y k_z \right\} + cp \right) +
i C_2 \left( T_x \epsilon_{yz} + cp \right) \right] $ \\
$H^{cs}$ & $ -\frac{1}{\sqrt{3}} \left[ P \left( k_x \rho_x + cp \right) + i B \left( \rho_x \left\{ k_y k_z \right\} + cp \right) +
i C_2 \left( \rho_x \epsilon_{yz} + cp \right) \right] $ \\
$H^{vs}$ & $H^{vs}_k + H^{vs}_\epsilon + H^{vs}_{\epsilon k}$ \\
$H^{vs}_k$ & $ -\frac{\hbar^2}{2 m} \left[ -3 \gamma_2 \left( U_{xx} k_x^2 + cp \right) - 6 \gamma_3 \left( U_{xy} \left\{ k_x k_y \right\} +
cp \right) \right] - \frac{e \hbar}{m c} \frac{3}{2} \left( U_x H_x + cp \right) $ \\
$H^{vs}_\epsilon$ & $ 2 D_u \left( U_{xx} \epsilon_{xx} + cp \right) + 2 D_u' \left( 2 U_{xy} \epsilon_{xy} + cp \right) $ \\
$H^{vs}_{\epsilon k}$ & $ \frac{3}{2_{\parbox[b][2mm][b]{1pt}{} }} \left[ C_4 \left( \epsilon_{yy} - \epsilon_{zz} \right) k_x + C_5' \left( \epsilon_{xy} k_y - \epsilon_{xz} k_z
\right) \right] U_x + cp $ \\
\hline \hline 
\multicolumn{2}{l}{ $cp$ means cyclic permutation, $\left\{ A B \right\}= \frac{1}{2} ( AB + BA )$,
$\tr \overline{\overline{\epsilon}}= \epsilon_{xx}+\epsilon_{yy}+\epsilon_{zz}$ }
\end{tabular}
\caption[Matrix elements of the 8-band \kp\ Hamiltonian]{Matrix elements of the 8-band \kp\ Hamiltonian.}
\label{tab:8bandH_elems}
\setlength{\extrarowheight}{0pt}
\end{table*}

\section{Conclusion}
\label{sec:k3_summary}

We have presented an implementation of an 8-band Finite Difference Effective Mass Approximation (EMA) method for calculating band structures. This implementation is faithful to the $T_d$ microscopic symmetry of bulk zincblendes. We show that this allows the heterostructure Hamiltonian to reproduce the $D_{2d}$ or the $C_{2v}$ point group symmetry of [001] grown zincblende heterostructures. As a consequence, all symmetry effects close to the zone center, including the spin splitting of the subbands, are correctly described. A simple expression for the $k^3$ splitting coefficient $\gamma_c$ is given in terms of the \kp\ parameters. We find an extra contribution to $\gamma_c$ that can amount to a few percent of its value. When the method is applied to symmetric heterostructures, linear splittings in $k$ are predicted as a consequence of the reduced symmetry. This is not the case for standard EMA implementations. The bands of asymmetric heterostructures are also studied and described in the context of the BIA model Hamiltonian for heterostructures. It is seen that, in the case studied, the SIA and BIA contributions to the spin splitting are of the same order of magnitude ($\alpha_R=40.3\times 10^{-10}$ eV$\cdot$cm and $\alpha_{\text{BIA}}=15.0\times 10^{-10}$ eV$\cdot$cm respectively) even though the well is composed by narrow gap materials. Therefore, an accurate description of the bands will require the inclusion of both effects.

\acknowledgments
The authors would like to thank D. L. Smith and J. N. Schulman for helpful discussions. This work has been supported by a subcontract with HRL, LLC in conjunction with the Office of Naval Research under contract No. MDA972-01-C-0002. A part of the work described in this paper was carried out at the Jet Propulsion Laboratory, California Institute of Technology, and was sponsored by the Defense Advanced Research Projects Agency's Spins in Semiconductors (SpinS) program.

\begin{table}[t]
\centering
\begin{ruledtabular}
\begin{tabular}{cc}
$O$ & $T_d$ \\
\hline
$\ket{\Gamma_8, -3/2}$ & $\ket{\Gamma_8, +1/2}$ \\
$\ket{\Gamma_8, -1/2}$ & $\ket{\Gamma_8, +3/2}$ \\
$\ket{\Gamma_8, +1/2}$ & $\ket{\Gamma_8, -3/2}$ \\
$\ket{\Gamma_8, +3/2}$ & $\ket{\Gamma_8, -1/2}$ \\
\end{tabular}
\end{ruledtabular}
\caption[Equivalence of basis functions for $O$ and $T_d$]{Correspondence in basis functions for $O$ and $T_d$ in Table 83 of Ref.~\onlinecite{KosterDimmockWheeler1963}.}
\label{tab:equivOTd}
\end{table}

\appendix

\section{Explicit form of the 8-band \kp\ Hamiltonian}
\label{app:8bandH}

The 8-band \kp\ Hamiltonian derived by Trebin \etal~\onlinecite{TrebinRosslerRanvaud1979} has been constructed using the theory of invariants~\cite{BirPikus}, and thus it correctly describes the $T_d$ symmetry of bulk zincblendes. In particular, it includes terms breaking the spin degeneracy of the bands at a general point in the Brillouin zone, making it ideal for the study of inversion asymmetry effects. It also accounts for the effects of strain and an external magnetic field.

Although the Hamiltonian is explicitly shown in Ref.~\onlinecite{TrebinRosslerRanvaud1979}, it is rewritten here for completeness and to correct two typographical errors. In order to minimize the probability of introducing hard-to-detect errors in the definition of the matrix elements, the Hamiltonian was first entered in Mathematica~\cite{Mathematica}. From there, the appropriate C code for each matrix element was generated automatically with the instruction \verb!CForm!. This method also has the advantage that it allows the algebraic operation of the Hamiltonian to find analytical forms for the dispersion relation very near the zone center (cf. Sec.~\ref{ssec:k3_Energy_expansion}).

The 8-band \kp\ Hamiltonian will be expressed in the basis $\left\{ \ket{\Gamma_6, +\frac{1}{2}} \right.$, $\ket{\Gamma_6, -\frac{1}{2}}$, $\ket{\Gamma_8, +\frac{3}{2}}$, $\ket{\Gamma_8, +\frac{1}{2}}$, $\ket{\Gamma_8, -\frac{1}{2}}$, $\ket{\Gamma_8, -\frac{3}{2}}$, $\ket{\Gamma_7, +\frac{1}{2}}$, $\left. \ket{\Gamma_7, -\frac{1}{2}} \right\}$. It can be written in a block diagonal form
\begin{equation}
H=
\begin{pmatrix}
H^{cc} & H^{cv} & H^{cs} \\
H^{vc} & H^{vv} & H^{vs} \\
H^{sc} & H^{sv} & H^{ss}
\end{pmatrix},
\label{eq:8bandHAM}
\end{equation}
where, of course, $\left( H^{\alpha \beta} \right)^\dagger = H^{\beta \alpha}$, $c$ refers to the two conduction band (CB) states, $v$ to the four heavy hole (HH) and light hole (LH) states and $s$ to the two spin-orbit split off (SO) states.

The constituent blocks of the Hamiltonian are shown in Table~\ref{tab:8bandH_elems}. The phases of the wavefunctions and the prefactors in Table~\ref{tab:8bandH_elems} are chosen in a way that all the parameters are real. The $\bm{\sigma}$ and the $\bm{\rho}$ matrices are the Pauli matrices; the $T$ matrices are given by
\begin{widetext}
\begin{alignat}{3}
T_x & = {\textstyle \frac{1}{3\sqrt{2}} } \left(
\begin{smallmatrix}
- \sqrt{3} & 0 & 1 & 0 \\
0 & -1 & 0 & \sqrt{3} \\
\end{smallmatrix} \right) \qquad &
T_y & = {\textstyle \frac{-i}{3\sqrt{2}} } \left(
\begin{smallmatrix}
\sqrt{3} & 0 & 1 & 0 \\
0 & 1 & 0 & \sqrt{3} \\
\end{smallmatrix} \right) \qquad &
T_z & = {\textstyle \frac{\sqrt{2}}{3} } \left(
\begin{smallmatrix}
0 & 1 & 0 & 0 \\
0 & 0 & 1 & 0 \\
\end{smallmatrix} \right) \notag \\
T_{xx} & = {\textstyle \frac{1}{3\sqrt{2}} } \left(
\begin{smallmatrix}
0 & -1 & 0 & \sqrt{3} \\
- \sqrt{3} & 0 & 1 & 0 \\
\end{smallmatrix} \right) \qquad &
T_{yy} & = {\textstyle \frac{1}{3\sqrt{2}} } \left(
\begin{smallmatrix}
0 & -1 & 0 & -\sqrt{3} \\
\sqrt{3} & 0 & 1 & 0 \\
\end{smallmatrix} \right) \qquad &
T_{zz} & = {\textstyle \frac{\sqrt{2}}{3} } \left(
\begin{smallmatrix}
0 & 1 & 0 & 0 \\
0 & 0 & -1 & 0 \\
\end{smallmatrix} \right) \notag \\
T_{yz} & = {\textstyle \frac{i}{2\sqrt{6}} } \left(
\begin{smallmatrix}
-1 & 0 & - \sqrt{3} & 0 \\
0 & \sqrt{3} & 0 & 1 \\
\end{smallmatrix} \right) \qquad &
T_{zx} & = {\textstyle \frac{1}{2\sqrt{6}} } \left(
\begin{smallmatrix}
-1 & 0 & \sqrt{3} & 0 \\
0 & \sqrt{3} & 0 & -1 \\
\end{smallmatrix} \right) \qquad &
T_{xy} & = {\textstyle \frac{i}{\sqrt{6}} } \left(
\begin{smallmatrix}
0 & 0 & 0 & -1 \\
-1 & 0 & 0 & 0 \\
\end{smallmatrix} \right).
\end{alignat}
\end{widetext}

The matrices $U$ are simply given by $U_i = T_i^{\dagger}$.

Note that a typographical error in the matrix $T_{xx}$ in Table~I of the original article by Trebin \etal~\cite{TrebinRosslerRanvaud1979} has been corrected here. The last equation in the group~(A3) of Ref.~\onlinecite{TrebinRosslerRanvaud1979} must also be corrected:
\begin{equation}
\label{eq:Xcorr}
X_{12}=-i \left( X_2^{(2)} - X_{-2}^{(2)} \right) /2
\end{equation}

\section{Basis state labels in the KDWS tables}

There is another remark about a point that can lead to confusion. Koster {\it et al.}~\cite{KosterDimmockWheeler1963} have developed a set of tables (KDWS tables) for the Clebsch-Gordan coefficients for point groups that are very helpful when constructing explicit subspace-invariant matrices or when checking the symmetry properties of the Hamiltonian. However, in their Table 83 the values they show can be used as displayed for the $O$ point group, but for $T_d$ the values should be taken according to the lookup table for the basis state labels shown in Table~\ref{tab:equivOTd}.

\section{Finite Difference Method solution of the Effective Mass Approximation equations}
\label{sec:k3_FDM}

In this appendix we describe the computational method used to find the heterostructure bands and eigenstates. Following the Effective Mass Approximation (EMA) theory~\cite{Bastard1986}, the 8 band \kp\ Hamiltonian of \refeq{eq:8bandHAM} is transformed into a set of eight linear, second order, ordinary differential equations. The appropriate boundary conditions are enforced and the equations are solved by means of a finite difference scheme.

\subsection{EMA Hamiltonian}
\label{ssec:k3_EMA_Ham}

Starting from the usual~\cite{LuttingerKohn1955,Bastard1981,Bastard1982} expansion of the wavefunction $\Psi_{{\bf k}_{\parallel}} \left( {\bf r} \right)$ as a linear combination of zone center Bloch states $u_{n0}$ modulated by a position dependent envelope $F_n (z)$ ($n$ is the band index):
\begin{equation}
\Psi_{{\bf k}_{\parallel}} \left( {\bf r} \right) = \sum_{n} e^{i {\bf k}_{\parallel} \cdot {\bf r} } F_n (z) \: u_{n0} \left( {\bf r} \right) ,
\end{equation}
the EMA prescription states that if $\Psi_{{\bf k}_{\parallel}} \left( {\bf r} \right)$ is the solution of
\begin{equation}
\left[ H \left( {\bf k}_{\parallel}, k_z; z \right) + U \left( z \right) \right] \Psi_{{\bf k}_{\parallel}} \left( {\bf r} \right) =
E \: \Psi_{{\bf k}_{\parallel}} \left( {\bf r} \right) ,
\end{equation}
then $F_n (z)$ will be the solution to
\begin{equation}
\left[ H \left( {\bf k}_{\parallel}, -i \partial_z; z \right) + U \left( z \right) \right] {\bf F} (z)  =
E \: {\bf F} (z) ,
\label{eq:env_func_ode}
\end{equation}
where ${\bf k}_{\parallel} = k_x \hat{x} + k_y \hat{y}$ is the electron wavevector in the $k_x-k_y$ plane, $U \left( z \right)$ is an external potential that varies only in $z$ (the change of material as a function of $z$ is included in $H \left( {\bf k}_{\parallel}, -i \partial_z; z \right)$) and ${\bf F} (z)$ is a multicomponent vector constructed from the different $F_n (z)$'s. In an 8-band theory, ${\bf F}$ would have 8 components, each one multiplying the conduction band (CB), heavy hole (HH), light hole (LH) and split off (SO) basis states.

On the other hand, the bulk \kp\ Hamiltonian can be expanded into its polynomial form for $k_z$ in the following manner:
\begin{equation}
{\bf H} \left( {\bf k} \right) = {\bf H}^{(2)} k^2_z + {\bf H}^{(1)} \left( {\bf k_{\parallel}} \right) k_z +
{\bf H}^{(0)} \left( {\bf k_{\parallel}} \right)
\label{eq:pol_exp_Ham}
\end{equation}

Putting together \refeq{eq:env_func_ode} and \refeq{eq:pol_exp_Ham}, we must solve the following system of coupled differential equations to obtain the energies and eigenstates of the system:
\begin{multline}
\left[ - {\bf H}^{(2)} \partial^2_z - i {\bf H}^{(1)} \left( {\bf k_{\parallel}} \right) \partial_z +
{\bf H}^{(0)} \left( {\bf k_{\parallel}} \right) + U(z) \right] {\bf F} (z)  =  \\
E \: {\bf F} (z).
\label{eq:F_ode}
\end{multline}

\subsection{The Finite Difference Method}
\label{ssec:k3_FDM}

There are several methods to solve numerically the system of coupled ordinary differential equations given by \refeq{eq:F_ode}, such as the transfer-matrix method~\cite{Kane1969,Chuang1991}, the finite element method~\cite{NakamuraShimizuKoshiba1991}, the basis expansion method~\cite{BauerAndo1988}, etc. In this study, the finite difference method (FDM) has been employed because of its conceptual simplicity, the ease of introduction of arbitrary fields, its ability to describe tunneling phenomena with only a few changes~\cite{LiuTingMcgill1996} and its numerical stability with respect to the transfer-matrix method, which requires the truncation of growing exponential states~\cite{KoInkson1988}.

\begin{figure}[b]
\centering
\epsfig{file=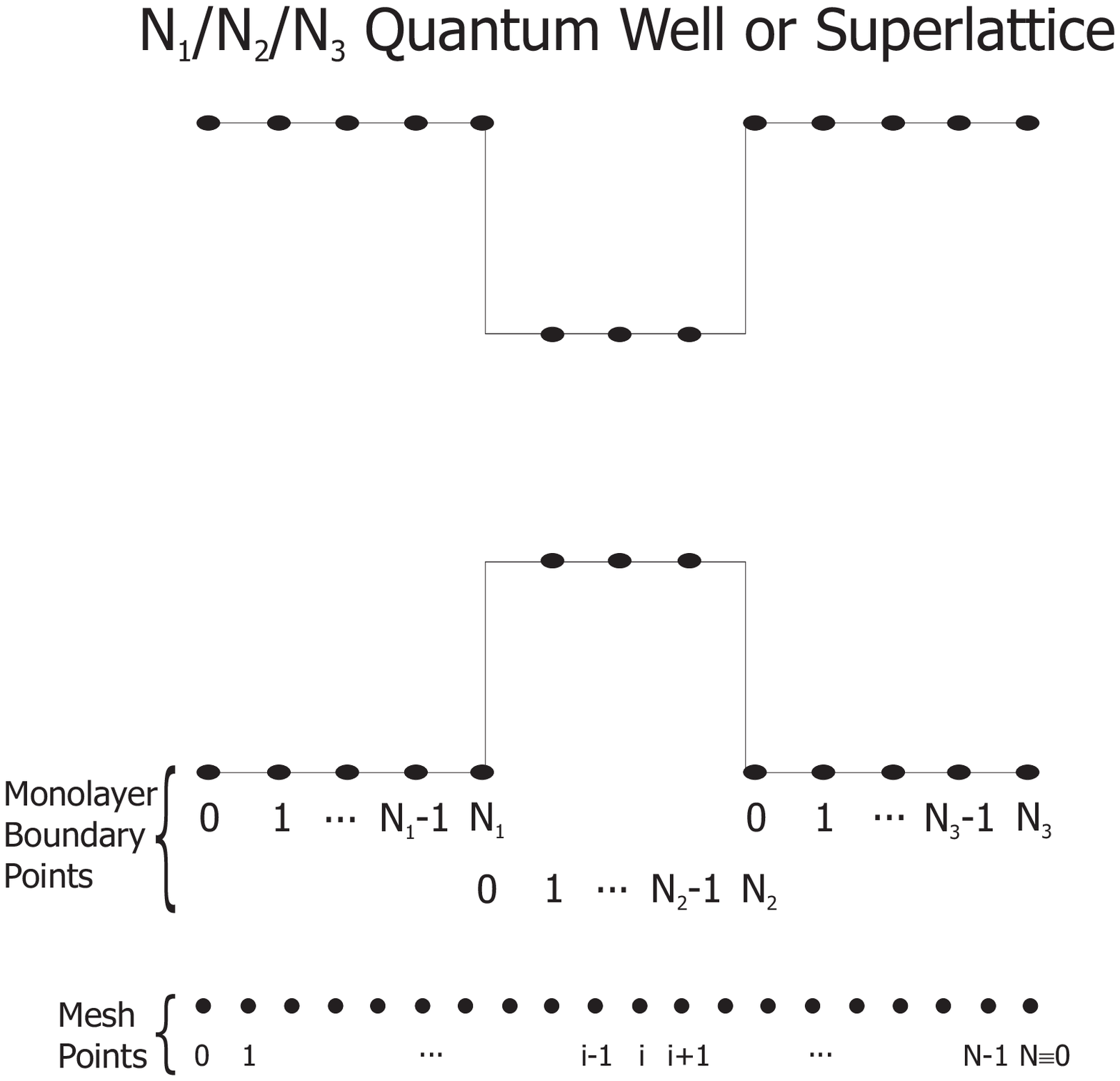, width=0.9\linewidth, clip=}
\caption[Mesh used in the solution of the EMA.]{Schematic of the structure under study, with points separating the monolayers, and mesh used when solving the Effective Mass Approximation equations. The mesh points need not coincide with the monolayer boundaries.}
\label{fig:discrete_mesh}
\end{figure}

In the finite difference method, the differential operators are first written in a Hermitian form and then substituted by their finite difference approximations over a discrete mesh (see Fig.~\ref{fig:discrete_mesh}) with $N$ points. Following Chuang and Chang~\cite{ChuangChang1997}, the following discretization scheme is used:
\begin{widetext}
\begin{multline}
\left. {\bf H}^{(2)}(z) \: \partial^2_z f \right|_{z_i} \rightarrow
\left. \partial_z \left( {\bf H}^{(2)}(z) \: \partial_z f \right) \right|_{z_i} \approx 
\frac{{\bf H}^{(2)}(z_{i+1})+{\bf H}^{(2)}(z_{i})}{2(\Delta z)^2} f(z_{i+1}) - \\ 
\frac{{\bf H}^{(2)}(z_{i+1})+2{\bf H}^{(2)}(z_{i})+{\bf H}^{(2)}(z_{i-1})}{2(\Delta z)^2} f(z_{i}) +
\frac{{\bf H}^{(2)}(z_{i-1})+{\bf H}^{(2)}(z_{i})}{2(\Delta z)^2} f(z_{i-1})
\label{eq:fd_H2}
\end{multline}
\begin{multline}
\left. - i {\bf H}^{(1)}(z) \: \partial_z f \right|_{z_i} \rightarrow
\frac{-i}{2} \left. \left[ {\bf H}^{(1)}(z) \: \partial_z f + \partial_z {\bf H}^{(1)}(z) f \right] \right|_{z_i} \approx \\
-i \frac{{\bf H}^{(1)}(z_{i+1})+{\bf H}^{(1)}(z_{i})}{4\Delta z} f(z_{i+1}) +
 i \frac{{\bf H}^{(1)}(z_{i-1})+{\bf H}^{(1)}(z_{i})}{4\Delta z} f(z_{i-1})
\label{eq:fd_H1}
\end{multline}
\end{widetext}
where $\Delta z$ is the separation between the mesh points, and $z_i$ is the position of the $i$-th mesh point.

Now, the application of the above equations to \refeq{eq:F_ode} yields the following system of $N$ algebraic equations:
\begin{equation}
{\bf H}_{i,i-1} {\bf F}_{i-1} + {\bf H}_{i,i} {\bf F}_{i} + {\bf H}_{i,i+1} {\bf F}_{i+1} = E {\bf F}_{i}
\label{eq:fd_eqs}
\end{equation}
where ${\bf F}_{i}$ is the eight-vector containing the envelope function components corresponding to the $i$-th mesh point. This eigenproblem can be written in matrix form to better appreciate its sparse structure:
\begin{widetext}
\begin{equation}
\begin{pmatrix}
{\bf H}_{0,0} & {\bf H}_{0,1} & {\bf 0} & \ldots & \ldots & \ldots & {\bf H}_{0,-1} \\
{\bf H}_{1,0} & {\bf H}_{1,1} & {\bf H}_{1,2} & {\bf 0} & \ldots & \ldots & {\bf 0} \\
{\bf 0} & {\bf H}_{2,1} & {\bf H}_{2,2} & {\bf H}_{2,3} & {\bf 0} & \ldots & {\bf 0} \\
\vdots & \ldots & \ddots & \ddots & \ddots & \vdots & \vdots \\
{\bf 0} & \ldots & \ldots & {\bf 0} & {\bf H}_{N-2,N-3} & {\bf H}_{N-2,N-2} & {\bf H}_{N-2,N-1} \\
{\bf H}_{N-1,N} & {\bf 0} & \ldots & \ldots & {\bf 0} & {\bf H}_{N-1,N-2} & {\bf H}_{N-1,N-1}
\end{pmatrix}
{\bf F} = E \: {\bf F}
\label{eq:eigensystem}
\end{equation}
\end{widetext}
where ${\bf F}$ is a column vector composed of the different ${\bf F}_i$'s.

The resulting discretized $8\times 8$ Hamiltonian matrices will be valid for both inner and interface mesh points without further modification:
\begin{align}
{\bf H}_{i,i} &= \frac{{\bf H}^{(2)}_{i+1}+2{\bf H}^{(2)}_{i}+{\bf H}^{(2)}_{i-1} }{2(\Delta z)^2} + {\bf H}^{(0)}_i + U_i 
\label{eq:Hii} \\
{\bf H}_{i,i+1} &= -\frac{ {\bf H}^{(2)}_{i+1}+{\bf H}^{(2)}_{i} }{2(\Delta z)^2} - i \frac{{\bf H}^{(1)}_{i+1}+{\bf H}^{(1)}_{i}}{4\Delta z} \\
{\bf H}_{i,i-1} &= -\frac{ {\bf H}^{(2)}_{i-1}+{\bf H}^{(2)}_{i} }{2(\Delta z)^2} + i \frac{{\bf H}^{(1)}_{i-1}+{\bf H}^{(1)}_{i}}{4\Delta z}
\label{eq:Himi}
\end{align}

${\bf H}_{0,-1}$ and ${\bf H}_{N-1,N}$ in \refeq{eq:eigensystem} express the boundary conditions (BCs) of the problem. When studying a quantum well, the BCs are that the wavefunction must vanish far from the well region. This is accomplished by setting the barrier region wide enough, and requesting
\begin{equation}
{\bf F}_{-1} = {\bf F}_N = 0
\end{equation}
which translates into
\begin{equation}
{\bf H}_{0,-1} = {\bf H}_{N-1,N} = {\bf 0}
\end{equation}

When finding the energies and states of a superlattice, the Bloch BCs apply, and the envelope function is requested to have the supercell periodicity $d$, modulated by a phase:
\begin{alignat}{3}
{\bf F}_N &= e^{i q d} {\bf F}_0 \qquad &\Rightarrow \qquad {\bf H}_{N-1,N} &= e^{i q d} {\bf H}_{N-2,N-1} \\
{\bf F}_{-1} &= e^{-i q d} {\bf F}_{N-1} \qquad &\Rightarrow \qquad {\bf H}_{0,-1} &= e^{-i q d} {\bf H}_{1,0}
\end{alignat}
where $q$ is the electron wavevector along the $z$ direction, and it has been assumed that the same material is at mesh points 0 and $N-1$.

\NewBibliography

\end{document}